\documentclass[12pt,a4paper]{article}%

\usepackage{amsfonts,color}
\usepackage[dvipsnames]{xcolor}
\usepackage{amsthm}
\usepackage{fancyhdr}
\usepackage{comment}

\usepackage{subcaption}
\usepackage{hyperref}
\usepackage[a4paper, top=1in, bottom=1in, left=1in, right=1in]%
{geometry}
\usepackage{times}
\usepackage{caption}
\usepackage{lscape}
\usepackage{amsmath}
\usepackage{spverbatim}
\usepackage{changepage}
\usepackage{amssymb}
\usepackage{graphicx}

\usepackage{placeins}
\usepackage{caption}
\usepackage{graphicx}%
\usepackage{mathptmx}
\usepackage{listings}
\usepackage{geometry}
\usepackage{spverbatim}
\usepackage[T1]{fontenc}
\usepackage[utf8]{inputenc}
\usepackage{authblk}
\usepackage{arydshln}
\usepackage{multirow}
\usepackage{authblk}

\newcommand{\bs}[1]{\pmb{#1}} 
\newcommand{\bb}[1]{\textbf{#1}} 
\newcommand{\T}{\intercal}
\newcommand{\tab}{\hspace*{0.75 cm}}

\usepackage{setspace}
\makeatletter
\renewcommand*{\@fnsymbol}[1]{\ifcase#1\or*\else **\fi}
\makeatletter

\usepackage{cite}
\renewcommand\@biblabel[1]{#1.}
\makeatletter
\def\@cite#1#2{$^{\mbox{\scriptsize #1\if@tempswa , #2\fi}}$}
\makeatother

\title{A Robust and Unified Framework for Estimating Heritability in Twin Studies using Generalized Estimating Equations}

\author[1]{Jaron Arbet\thanks{jaron.arbet@ucdenver.edu}}
\author[2]{Matt McGue}
\author[3]{Saonli Basu\thanks{saonli@umn.edu}}
\affil[1]{Department of Biostatistics and Informatics, University of Colorado Anschutz Medical Campus, Aurora, CO}
\affil[2]{Department of Psychology, University of Minnesota, Minneapolis, MN}
\affil[3]{Department of Biostatistics, University of Minnesota, Minneapolis, MN}

\let\oldabstract\abstract
\let\oldendabstract\endabstract
\makeatletter
\renewenvironment{abstract}
{%
	{\list{}{\addtolength{\leftmargin}{-1em}
			\listparindent 1.5em%
			\itemindent    \listparindent%
			\rightmargin   \leftmargin%
			\parsep        \z@ \@plus\p@}%
		\item\relax}%
	{\endlist}%
	\oldabstract}
{\oldendabstract}
\makeatother

\begin{document}
	\maketitle
\begin{abstract}
	The development of a complex disease is an intricate interplay of genetic and environmental factors. The ‘heritability’ of a quantitative trait
measures the proportion of total trait variance due to genetic factors in a given population.  Studies with monozygotic (MZ) and dizygotic (DZ) twins allow us to estimate heritability by fitting an ``ACE'' model which estimates the proportion of trait variance explained by additive genetic ($A$), common  shared environment ($C$), and non-shared environmental ($E$) latent effects, thus helping us better understand disease risk and etiology.  In this paper, we develop a flexible generalized estimating equations framework (``GEE2'') for fitting twin ACE models that requires minimal distributional assumptions; only the first two moments need to be correctly specified.  We show that two commonly used methods for estimating heritability, the normal ACE model (``NACE'') and Falconer's method, can both be fit within this unified GEE2 framework, which additionally provides robust standard errors.  Although the traditional Falconer's method cannot directly adjust for covariates, the corresponding GEE2 version (``GEE2-Falconer'') can incorporate both mean and variance-level covariate effects (e.g. let heritability vary by sex or age).  Given non-normal data, we show that the GEE2 models attain significantly better coverage of the true heritability compared to the traditional NACE and Falconer's methods. Finally, we demonstrate an important scenario where the NACE model produces biased estimates of heritability while Falconer’s method remains unbiased. Overall, we recommend using the robust and flexible GEE2-Falconer model for estimating heritability in twin studies.
\end{abstract}
\clearpage
\section{Introduction}
\label{sec_intro}
\tab Twins and family studies have proven to be powerful instruments for understanding the inheritance of complex phenotypes\cite{neale2013twinsbook}. The `inheritance' or `heritability' of a quantitative trait measures the proportion of total trait variance due to genetic factors in a given population.   The accurate estimation and inference of heritability is often of primary interest as it gives us some basic understanding of disease risk and etiology. For a review of various concepts and methods for estimating heritability, see\cite{visscher2008heritability,tenesa2013heritability}.  In this paper, we focus on the twin ACE model\cite{neale2013twinsbook, rijsdijk2002analytic, polderman2015meta} which compares the resemblance among monozygotic (MZ) and dizygotic (DZ) twins in order to estimate heritability.
Specifically, the trait variance of each twin pair is partitioned into additive genetic ($A$), common shared family environment ($C$), and non-shared environmental ($E$) variance components.  The parameters of this twin ACE model are estimated using simple method of moment estimators called ``Falconer's equations''\cite{rijsdijk2002analytic, polderman2015meta, falconer1975introduction}; structural equation models (SEM)\cite{neale2013twinsbook,rijsdijk2002analytic}, or  likelihood based approaches assuming normality of the trait (henceforth referred to as the ``normal ACE model'' or ``NACE'')\cite{rabe2008biometrical,wang2011statistical,feng2009analysis,mcardle2005mixed}. A recent comprehensive meta-analysis reported heritability estimates for 17,804 traits based on the past 50 years of twin studies\cite{polderman2015meta}. This meta analysis\cite{polderman2015meta} reported heritability estimates from multiple twin studies using both the normal ACE model and Falconer's method, and showed substantial differences in the reported estimates (see their Supp. Figures 9-10 and Supp. Section 5.7).   In this paper, we illustrate how failure to satisfy certain model assumptions could potentially cause substantial differences in the heritability estimates reported by these two methods. This is particularly useful as overestimation of heritability through twin-studies are often attributed to a potential reason behind `missing heritability'~\cite{maher2008personal,manolio2009finding,zuk2012mystery}.

The normal ACE model (NACE) is a popular approach for estimating heritability in twin studies\cite{rabe2008biometrical,wang2011statistical,feng2009analysis,mcardle2005mixed}.  However, the NACE model assumes the trait is normally distributed, and results in Section \ref{sec_results} demonstrate that when the assumption of normality is violated, the NACE model can lead to poor coverage of the true heritability parameter. Moreover, the NACE model assumes the ACE variance parameters are equal for both MZ and DZ twin types. We show that the NACE produces biased estimates of heritability given deviations from this assumption. Alternatively, one could use Falconer's distribution-free method of moment estimators\cite{falconer1975introduction,rijsdijk2002analytic}.  Unlike the traditional NACE model, Falconer's method allows the total variance to differ between MZs and DZs and only assumes the \textit{proportion} of total variance explained by genetic and environmental effects to be the same for MZs and DZs.  In doing so, Falconer's method makes less stringent assumptions about the twin population.  In particular, we demonstrate that Falconer's method can generate valid estimates of heritability when the ACE variance parameters differ between MZ and DZ twins; while the traditional NACE model generates biased estimates in such settings.

Researchers are often interested  in estimating heritability for highly non-normal traits such as binary case/control data, discrete counts, and skewed or heavy-tailed continuous data.  Moreover, often the trait of interest doesn't appear to follow any standard parametric distribution (see Figure \ref{fig_RDAhist} for examples).  Existing approaches to estimating heritability for non-normal traits include generalized linear mixed effect models \cite{jamsen2013specification,burton_binary,scurrah_poisson}.  Recently Kirkpatrick and Neale\cite{kirkpatrick2016applying} developed three parametric models for estimating ACE variance components in count phenotypes. However, in practice, the estimation and inferences from these models may be sensitive to departures from the parametric distributional assumptions.  In addition, often one will fit several different parametric models and then use model selection criteria to pick the ``best fitting'' parametric model. This may lead to biased results if the model selection procedure is not accounted for while conducting inferences\cite{polderman2015meta}.  Thus a more flexible semi-parametric (or non-parametric) approach to estimating heritability may be desirable for non-normally distributed outcomes.

In this paper, we propose a robust, unified framework for estimating heritability in twin studies using second-order generalized estimating equations (``GEE2'').
The semi-parametric GEE2 models require only the first two moments to be correctly specified, and thus can be used to estimate heritability in a wide variety of phenotypes, without explicitly modeling the underlying true parametric distribution.  We show that two traditional methods for estimating heritability (NACE and Falconer's method) can both be fit within the GEE2 framework, which additionally provides robust standard errors.  Although
the traditional Falconer's method cannot directly adjust for covariates, we show that the corresponding
GEE2 version (`GEE2-Falconer') can accommodate  covariate effects for both mean
and variance-level parameters (e.g. let heritability vary by sex or age). Given a non-normal trait, we show that the robust GEE2 models produce significantly better coverage rates of the true heritability compared to the traditional NACE and Falconer's methods. Finally, we demonstrate that if the ACE variance parameters differ between MZ and DZ twins, then the NACE produces biased estimates of heritability, while Falconer's method remains unbiased under weaker assumptions and therefore should be recommended.  All methods are compared via simulations and with an application to the Minnesota Center for Twins and Family Study\cite{mcgue2013genome}.

\section{Methods}
\label{sec_methods}
\tab An outline of the Methods section is as follows: in Sections \ref{sec_normACE}-\ref{sec_falconer}, we review the traditional NACE and Falconer's method for estimating heritability in twin studies. Then in Section 2.3 we develop robust GEE2 versions of both models, and show how the GEE2 framework can allow heritability to vary as a function of covariates (e.g. sex or age).

For all methods, assume a study with $N_{MZ}$ and $N_{DZ}$ pairs of monozygotic and dizygotic twins, and let $N=N_{MZ}+N_{DZ}$ be the total number of twin pairs.  Let $\bs{y}_{z}=(y_{z_1}, y_{z_2})^\T$ be a quantitative response measured on both twins (1 and 2) for a given twin pair, with zygosity ``$z$'' equal to ``MZ" or ``DZ"; and $\bs{x}_{z}^\T$ is a $2\times P$ matrix of $P$ covariates for both twins.  Then the twin ACE model for a given pair of twins of type $z$ is defined as:
\begin{align}\label{eqn_generaltwinACE}
&\bs{y}_{z}= \bs{x}_z^\T\bs{\beta}+\bs{A}_{z}+\bs{C}_{z}+\bs{E}_{z},
\end{align}
 where $\mathbb{E}(\bs{y}_{z})=\bs{x}_{z}^\T\bs{\beta}$ and $\text{cov}(\bs{y}_z)=\bs{\Sigma}_z = \text{cov}(\bs{A}_{z}) + \text{cov}(\bs{C}_{z}) + \text{cov}(\bs{E}_{z})$.  The ACE random effects are defined to have the following mean and covariance structures: 
 \begin{align*}
 \bs{A}_{z} \sim \big(\bs{0}, \  \sigma^2_{A_z}\bs{\mathrm{K}}_z\big), \ \ \ \bs{C}_{z} \sim \big(\bs{0}, \  \sigma^2_{C_z}\bs{\mathrm{J}}\big), \ \ \ \bs{E}_{z} \sim \big(\bs{0}, \  \sigma^2_{E_z}\bs{\mathrm{I}}\big)
 \end{align*}
where $\bs{\mathrm{I}}$ is a $2\times 2$ identity matrix, $\bs{\mathrm{J}}$ is a $2\times 2$ matrix of ones, and $\bs{\mathrm{K}}_z=\begin{bmatrix} 1 & w_z \\ w_z & 1 \end{bmatrix}$ is the ``genomic relationship matrix''.  Note $w_z=1$ for MZ twins and $w_z=0.5$ for DZ twins, since MZ twins share 100\% of their genome while DZ twins share $50\%$ of their genome on average.  The parameters $\sigma_{A_{z}}^2$, $\sigma_{C_{z}}^2$, and $\sigma_{E_{z}}^2$  represent additive genetic, shared and non-shared variance parameters for twin type $z$. 
The primary interest for this `ACE' model is to estimate heritability, which is defined as the proportion of total trait variance due to additive genetic effects:
\begin{align*}
h^2=\frac{\sigma^2_{A_{MZ}}}{\sigma^2_{A_{MZ}}+\sigma^2_{C_{MZ}}+\sigma^2_{E_{MZ}}}=\frac{\sigma^2_{A_{DZ}}}{\sigma^2_{A_{DZ}}+\sigma^2_{C_{DZ}}+\sigma^2_{E_{DZ}}}
\end{align*}
Often we are also interested in estimating the proportion of trait variance due to shared environmental effects:
\begin{align*}
c^2=\frac{\sigma^2_{C_{MZ}}}{\sigma^2_{A_{MZ}}+\sigma^2_{C_{MZ}}+\sigma^2_{E_{MZ}}}=\frac{\sigma^2_{C_{DZ}}}{\sigma^2_{A_{DZ}}+\sigma^2_{C_{DZ}}+\sigma^2_{E_{DZ}}}
\end{align*}
Finally, the proportion of trait variance due to non-shared environmental effects is defined as: $e^2= 1-h^2-c^2.$ Note that equation (\ref{eqn_generaltwinACE}) allows distinct variance parameters for the different twin types ($z$ = ``MZ" or ``DZ"). However, all these distinct variance parameters are not estimable in a standard twin study with MZ and DZ twins. Hence the different methods to estimate heritability make certain assumptions about the underlying MZ, DZ populations to generate a valid identifiable model. Below we describe two such common approaches to estimate heritability. 

Without loss of generality, for the remainder of this paper we assume the response is centered such that $\mathbb{E}(\bs{y}_{z})=\bs{0}.$  Given that our primary focus in on variance parameters, fixing the mean equal to zero will greatly simplify formulas and thus help build intuitive connections between the various models considered in this paper.  However, in practice, both the NACE and GEE2 models described below can incorporate both mean and variance-level covariate effects.

\subsection{Normal ACE Model for Twin Studies}\label{sec_normACE}
The NACE model assumes the random effects are normally distributed such that $\bs{y}_{z}$ has the following log-likelihood function:
\begin{align}
\log\Bigl(f(\bs{y}_{z}|\bs{\alpha})\Bigr) = -0.5\big(\log(|\bs{\Sigma}_{z}|)\big)+\bs{y}_{z}^\T\bs{\Sigma}_{z}^{-1}\bs{y}_{z}+2\log(2\pi),\nonumber
\end{align} 
where $\bs{\Sigma}_{z}=\begin{bmatrix}
\sigma^2_A+\sigma^2_C+\sigma^2_E & w_{z}\sigma^2_A+\sigma^2_C \\ w_{z}\sigma^2_A+\sigma^2_C & \sigma^2_A+\sigma^2_C+\sigma^2_E
\end{bmatrix}$ and $\bs{\alpha}=(\sigma^2_A,\sigma^2_C,\sigma^2_E)$. The NACE makes a few simplifying assumptions to the model in equation (\ref{eqn_generaltwinACE}), such as $\sigma^2_{A_{z}}=\sigma^2_A$, $\sigma^2_{C_{z}}=\sigma^2_C$ and 
$\sigma^2_{E_{z}}=\sigma^2_E$.  Hence under the ``NACE" model, $ \text{cov}(\bs{y}_{z})=\bs{\Sigma}_{z}=\sigma^2_A \bb{K}_{z}+\sigma^2_C \bb{J}+\sigma^2_E \bb{I}$, for  $z$ = ``MZ'' or ``DZ''.  The parameters of interest are jointly estimated over the MZ and DZ families. See\cite{rabe2008biometrical,wang2011statistical,feng2009analysis,mcardle2005mixed} for a review of the popular normal ACE twin model (``NACE'').  

 For a given twin pair, the estimating equations for $\bs{\alpha}$ can be derived as:
\begin{align*} 
&\bs{u}(\bs{\alpha})_{NACE}=\frac{\partial}{\partial \bs{\alpha}}log f(\bs{y}_z|\bs{\alpha})= \Big(\frac{\partial}{ \partial \sigma^2_A}logf, \ \frac{\partial}{ \partial \sigma^2_C}logf, \ \frac{\partial}{ \partial \sigma^2_E}logf\Big)^\T=\bs{0}
\end{align*}
Assuming the multivariate-normal distribution $\log {f}(\bs{y}_z|\bs{\alpha})$ is correct, then under the regularity conditions of maximum likelihood estimation\cite{boos_springer}:
\begin{align}\label{eqn_normACE_AR}
& \sqrt{N}(\hat{\bs{\alpha}}-\bs{\alpha})\overset{D}{\to} MVN\big(\bs{0}, \bs{V}^{-1}\big), \tab \bs{V}=-\mathbb{E}\Big(\frac{\partial^2}{\partial \bs{\alpha} \partial \bs{\alpha}^\T}\log {f}(\bs{y}_z|\bs{\alpha})\Big),
\\ &
\hat{Cov}(\bs{\hat{\alpha}})=\frac{1}{N}\bs{\hat{V}}^{-1}= \frac{1}{N}\Big[\frac{-1}{N}\sum_{1}^N \frac{\partial^2}{\partial \bs{\alpha} \partial \bs{\alpha}^\T}logf(\bs{y}_z|\bs{\alpha})\Big]^{-1}_{\bs{\alpha}=\bs{\hat{\alpha}}} \nonumber
\end{align}
where the summation in $\hat{Cov}(\hat{\bs{\alpha}})$ is taken with respect to all $N$ twin pairs.
After obtaining $\hat{\bs{\alpha}}$ and $\hat{Cov}(\hat{\bs{\alpha}})$, we used the Delta-Method to construct approximate Wald tests and $95\%$ confidence intervals for $h^2$ and $c^2$ \big(e.g. $\hat{h}^2\pm1.96\hat{SE}(\hat{h}^2)$\big).  It is worth noting that if the assumed multivariate-normal likelihood function is misspecified (as is often the case in practice), then in general, equation (\ref{eqn_normACE_AR}) will not hold. Finally, we used the \textit{twinlm()} function from the \textit{mets} R package\cite{mets} to implement the NACE model.

\subsection{Falconer's Method of Moment Estimators}\label{sec_falconer}

``Falconer's equations'' use method of moments to estimate heritability in twin studies\cite{falconer1975introduction,rijsdijk2002analytic}.  Falconer's estimators for $h^2$ and $c^2$ are defined as:
\begin{align}\label{eqn_falc}
\hat{h}^2_{Falc} = 2\big(r_{MZ}-r_{DZ}\big), \ \hat{c}^2_{Falc}=2r_{DZ}-r_{MZ},
\end{align}
where $r_{MZ}$ and $r_{DZ}$ are Pearson's sample correlation coefficients for the MZ and DZ twins respectively.  Following the notation of equation (\ref{eqn_generaltwinACE}), Falconer's estimators are derived as follows:
\begin{align*}
& \rho_{MZ}=Corr(\bs{y}_{MZ_1},\bs{y}_{MZ_2})=\frac{Cov_{MZ}}{Var(\bs{y}_{MZ})}=\frac{\sigma^2_{A_{MZ}}+\sigma^2_{C_{MZ}}}{\sigma^2_{A_{MZ}}+\sigma^2_{C_{MZ}}+\sigma^2_{E_{MZ}}} = h^2+c^2 \\
&\rho_{DZ}=Corr(\bs{y}_{DZ_1},y_{DZ_2})=\frac{Cov_{DZ}}{Var(\bs{y}_{DZ})}=\frac{0.5\sigma^2_{A_{DZ}}+\sigma^2_{C_{DZ}}}{\sigma^2_{A_{DZ}}+\sigma^2_{C_{DZ}}+\sigma^2_{E_{DZ}}} = 0.5h^2+c^2 \\
& \implies 2(\rho_{MZ}-\rho_{DZ})=h^2, \ \ 2\rho_{DZ}-\rho_{MZ}=c^2,
\end{align*}
where $\rho_{MZ}$ and $\rho_{DZ}$ are the population correlation coefficients between MZ and DZ twins respectively, and $Var(\bs{y}_{z})$ is the variance of both twins for a given zygosity type $z$.  Unlike the NACE, Falconer's method only requires the variance \textit{proportions} ($h^2, \ c^2, \ e^2$) to be equal for both MZ and DZ twins, but allows the magnitude of the ACE variance components ($\sigma^2_{A_z}, \sigma^2_{C_z},\sigma^2_{E_z}$) to differ between MZs and DZs.  In Section \ref{sec_NACEbiased}, we demonstrate that when the population variance differs between MZ and DZ twins (but the proportions $h^2, c^2, e^2$ are equal between twins), then NACE produces biased estimates of heritability while Falconer's method remains unbiased.

However, Falconer's approach is often criticized for being unable to directly adjust for covariates and there is no straightforward way to estimate the standard errors of the estimators. One could potentially derive the standard errors of the estimators based on asymptotic results of Pearson's sample correlation coefficient\cite{bowley1928standard}:
\begin{align*}
&\hat{SE}(\hat{h}^2_{Falc})\approx\sqrt{4\Big(\hat{Var}(r_{MZ})+\hat{Var}(r_{DZ})\Big)} = \sqrt{4\Big(\frac{(1-r_{MZ}^2)^2}{N_{MZ}}+\frac{(1-r_{DZ}^2)^2}{N_{DZ}}\Big)} \nonumber\\
&\hat{SE}(\hat{c}^2_{Falc})\approx\sqrt{4\hat{Var}(r_{DZ})+\hat{Var}(r_{MZ})} = \sqrt{4\Big(\frac{(1-r_{DZ}^2)^2}{N_{DZ}}\Big)+\frac{(1-r_{MZ}^2)^2}{N_{MZ}}} \nonumber
\end{align*}
Then using the estimated standard errors, we can construct approximate $95\%$ Wald-type confidence intervals for $h^2$ and $c^2$. However, we demonstrate through simulations that the aforementioned standard errors can produce poor coverage rates of the true heritability parameter. On the otherhand, our proposed GEE2-Falconer approach gives robust standard error estimates for the estimated heritability parameter. Additionally, although the traditional Falconer's method cannot adjust for covariate effects, we show that the GEE2 version of Falconer's method can incorporate covariate effects for both mean and variance-level parameters.

In the following section, we develop a unified framework for fitting both the NACE and Falconer's methods using a ``GEE2'' approach. Our proposed approach provides the flexibility to adjust for covariates (in both mean or variance-level parameters) and can accommodate inference of heritability parameter for non-normal traits by generating robust standard error estimates.
\subsection{GEE2 ACE Model for Twin Studies}\label{sec_GEE2ACE}
Liang and Zeger\cite{liang1986longitudinal} originally proposed the ``GEE1'' estimating equations which allow valid large-sample estimation and inferences on \textit{first} order moment parameters (e.g. mean-level parameters ``$\bs{\beta}$''), while allowing all higher-order moments to be misspecified. The essential assumption of GEE1 is that the trait is some member of the linear exponential family with only the first-moment structure required to be correctly specified, e.g. $\mathbb{E}(\bs{y_z})= \mathrm{\bs{x_z}}^\T\bs{\beta}$ \big(or $g^{-1}(\mathrm{\bs{x_z}}^\T\bs{\beta})$ if using a link function\big).

However, in applications where one is interested in conducting inference on both mean and variance-level parameters, GEE1 is no longer applicable. Prentice and Zhao\cite{prentice1991estimating} extended GEE1 by proposing the ``GEE2'' estimating equations which allow for valid inference on both mean \textit{and} variance level-parameters with minimal distributional assumptions.  The key assumption of GEE2 is that $\bs{y}$ is a member of the quadratic exponential family with the first two moments correctly specified \big(i.e. $\mathbb{E}(\bs{y_z})$ and $Cov(\bs{y_z})$\big); while all higher-order moments are allowed to be misspecified. If the aforementioned assumptions of GEE2 are satisfied, then GEE2 can consistently jointly estimate both mean-level parameters ($\bs{\beta}$) and variance-level parameters ($\bs{\alpha}$), as well as provide valid Wald tests and confidence intervals for all parameters.  For a complete review of GEE2, see\cite{prentice1991estimating,ziegler1998generalised,ziegler_GEEbook}. 
We show that both the NACE and Falconer's method can be fit within a unified GEE2 framework.  

\subsubsection{GEE2-NACE}\label{sec_GEE2_NACE}
We will first derive the NACE model under GEE2 framework, where we use the same notation and assumptions from Section \ref{sec_normACE} (e.g. assume the individual ACE variance component parameters are the same for both MZ and DZ twins).
Let the outcome for a given pair of twins $\bs{y}_z=(y_{z_1}, y_{z_2})$ be an arbitrary member of the quadratic exponential family with mean parameters ($\bs{\beta}$) and variance parameters ($\bs{\alpha}$):
\begin{align*} 
f(\bs{y}_z|\bs{\beta},\bs{\alpha})=\exp\Big\{h(\bs{\beta},\bs{\alpha})^\T\bs{y}_z+c(\bs{\beta}, \bs{\alpha})+d(\bs{y}_z)+\bs{y}_z^\T\bs{\mathrm{D}}(\bs{\beta},\bs{\alpha})\bs{y}_z\Big\}
\end{align*}
Without loss of generality, assume $\bs{\beta}=\bs{0}$ is fixed, and let $\bs{\alpha}=(\sigma^2_A, \ \sigma^2_C, \ \sigma^2_E)$ be the variance parameters.  Then define $\bs{\Gamma}_{z}$ and $\bs{\gamma}_{z}$ to be the population and sample variances in the following vectorized form $\bs{\Gamma}_{z} = (\sigma^2_A+\sigma^2_C+\sigma^2_E, \ \sigma^2_A+\sigma^2_C+\sigma^2_E, \ w_{z}\sigma^2_A+\sigma^2_C)^\T$ and $\bs{\gamma}_{z}=\big(y_{z_1}^2,\ y_{z_2}^2, \ y_{z_1}y_{z_2}\big)^\T$.
Define $ \bs{\mathrm{f}}_{z}= \bs{\gamma}_{z} - \bs{\Gamma}_{z}.$ Then Prentice and Zhao\cite{prentice1991estimating} derived the following estimating equations assuming $\bs{y}_{z}$ belongs to the quadratic exponential family:
\begin{align} \label{eqn_GEE2EE}
u_z(\bs{\alpha})= \bs{D}_{z}^\T \bs{\Omega}_{z}^{-1}\bs{\mathrm{f}}_{z}=\bs{0}, \ \ \text{ where }
\bs{D}_{z}=\begin{bmatrix}
\frac{ \partial \bs{\Gamma}_{z}}{\partial \bs{\alpha}^\T}
\end{bmatrix}, \ \ \  \bs{\Omega}_{z} = \begin{bmatrix}
Cov(\bs{\gamma}_{z})
\end{bmatrix}
\end{align}
Note that $\bs{\Omega}_z$ is the ``working covariance structure'' of the sample covariance vector $\bs{\gamma}_z$.  Recall from GEE2 theory that only $\mathbb{E}(\bs{y}_z)$ and $Cov(\bs{y}_z)=\bs{\Sigma}_z$ are required to be correctly specified, whereas the working covariance structure $\bs{\Omega}_z$ is allowed to be misspecified and one can still obtain valid inference for both mean and variance parameters ($\bs{\beta},\bs{\alpha})$ in large samples.  The ``normal working covariance''\cite{prentice1991estimating} for the GEE2-NACE model is defined as:
\begin{align*}
	& \bs{\Omega}_{z,Norm}= \scriptsize \begin{bmatrix}
	2(\sigma^2_A+\sigma^2_C+\sigma^2_E)^2 & 2(w_{z}\sigma^2_A+\sigma^2_C)^2 & 2(\sigma^2_A+\sigma^2_C+\sigma^2_E)(w_{z}\sigma^2_A+\sigma^2_C) \\
	2(w_{z}\sigma^2_A+\sigma^2_C)^2 & 2(\sigma^2_A+\sigma^2_C+\sigma^2_E)^2  & 2(\sigma^2_A+\sigma^2_C+\sigma^2_E)(w_{z}\sigma^2_A+\sigma^2_C) \\
	2(\sigma^2_A+\sigma^2_C+\sigma^2_E)(w_{z}\sigma^2_A+\sigma^2_C) & 2(\sigma^2_A+\sigma^2_C+\sigma^2_E)(w_{z}\sigma^2_A+\sigma^2_C) & (w_{z}\sigma^2_A+\sigma^2_C)^2+(\sigma^2_A+\sigma^2_C+\sigma^2_E)^2 
	\end{bmatrix}
	\end{align*}
Put simply, the normal working covariance assumes that \textit{all} moments of $\bs{y}_z$ follow a multivariate normal distribution.  Given an initial estimate $\bs{\alpha}_0$, a modified Newton-Raphson algorithm is used to iteratively update the estimator as follows\cite{prentice1991estimating}:
\begin{align} \label{eqn_updater}
\bs{\alpha}_u=
\bs{\alpha}_0
+ \Big\{\big( \sum_{1}^N \bs{D}_{z}^\T \bs{\Omega}_{z}^{-1} \bs{D}_{z}\big)^{-1}\big( \sum_{1}^N \bs{D}_{z}^\T \bs{\Omega}_{z}^{-1} \bs{\mathrm{f}}_{z}\big)\Big\}_{\alpha=\alpha_0}. 
\end{align}
Next, the following robust estimator for $Cov(\hat{\bs{\alpha}})$ is used\cite{prentice1991estimating}:
\begin{align} \label{eqn_est_covar}
& \hat{Cov}(\hat{\bs{\alpha}})=N^{-2}\bs{\Psi}^{-1}\Big(\sum_{1}^N \bs{D}_z^\T \bs{\Omega}_z^{-1}\bs{\mathrm{f}}_z \bs{\mathrm{f}}_z^\T\bs{\Omega}_z^{-1}\bs{D}_z\Big)\bs{\Psi}^{-1}\Big|_{\bs{\alpha}=\bs{\hat{\alpha}}} \ \  \text{ where } \  \bs{\Psi}= N^{-1}\sum_{1}^N \bs{D}_z^\T \bs{\Omega}_z^{-1}\bs{D}_z
\end{align}
Then robust standard errors for $\hat{\bs{\alpha}}$ can be obtained by taking the square-root of the diagonal of $\hat{Cov}(\bs{\hat{\alpha}})$.  Note that $\frac{1}{N}\bs{\Psi}^{-1}$ is the ``model-based'' variance of $\hat{\bs{\alpha}}$, derived from the implied likelihood function which follows the quadratic exponential family.  In general, this model-based variance estimator is incorrect when the implied likelihood function is misspecified.  The inside ``empirical-variance'' term $\Big(\sum_{1}^N \bs{D}_z^\T \bs{\Omega}_z^{-1}\bs{\mathrm{f}}_z \bs{\mathrm{f}}_z^\T\bs{\Omega}_z^{-1}\bs{D}_z\Big)$ is a consistent nonparametric estimator of the true variance of $\hat{\bs{\alpha}}$.  The reason these standard errors are ``robust'' is because although we allow $\bs{\Omega}_z=Cov(\bs{\gamma}_z)$ to be misspecified when estimating $\bs{\hat{\alpha}}$, the standard errors ``correct'' this by using a consistent nonparametric estimator of $Cov(\bs{\gamma}_z)$ through the inside-term $\bs{\mathrm{f}}_z\bs{\mathrm{f}}_{z}^\T=(\bs{\gamma}_z-\bs{\Gamma}_z)(\bs{\gamma}_z-\bs{\Gamma}_z)^\T.$  In contrast, the standard errors for the traditional NACE model are completely determined by the multivariate normal likelihood function, which if misspecified, can lead to poor coverage rates of the true variance parameters.

Note in Supplemental Material Section 2, the estimating equations for the NACE and GEE2-NACE models are derived and shown to be identical, thus both models will produce identical point estimates (with perhaps slight differences due to different software implementations).  However, we show through simulations that the GEE2-NACE model, which uses robust standard errors, provides a better coverage rate of the true heritability parameter given non-normal data.

Lastly, it is possible to allow the ACE variance components to differ as a function of covariates.  For example, suppose one wants to allow the ACE variance components to vary as a function of sex.  Then for a given twin pair, we can redefine the ACE variance components as follows:
\begin{align*}
g(\sigma^2_{A})= a_{0}+a_1Sex, \ \ \ g(\sigma^2_{C})= c_{0}+c_1Sex, \ \ \ g(\sigma^2_{E})= e_{0}+e_1Sex
\end{align*}
where $g(.)$ is a specified link function (e.g. identity or log-link), and $Sex$ represents the sex of a given twin pair.  Note that we assume both twins within a given pair have the same sex, thus we do not allow for the case of mixed-gender DZ twins.  Now our new variance parameters of interest are: $\bs{\alpha}=(a_0,a_1,c_0,c_1,e_0,e_1)$, and equations (\ref{eqn_updater}-\ref{eqn_est_covar}) can be used to obtain the estimates and standard errors.  Finally, the heritabilities for males and females are defined as:
\begin{align*}
h^2_{Male}=\frac{g^{-1}(a_0+a_1)}{g^{-1}(a_0+a_1)+g^{-1}(c_0+c_1)+g^{-1}(e_0+e_1)}, \ \ h^2_{Female}=\frac{g^{-1}(a_0)}{g^{-1}(a_0)+g^{-1}(c_0)+g^{-1}(e_0)}
\end{align*}
Note that $c^2$ and $e^2$ for males and females would be defined similarly.  The Delta method is used to obtain the final standard errors for $\hat{h}^2_{Male}, \hat{h}^2_{Female}$.  This framework can easily be extended to account for other covariate effects as long as the covariate takes on the same values within a given twin pair (e.g. age).  Accounting for ACE covariate effects with covariates that differ within a given twin pair is left for future work.

\subsubsection{GEE2-Falconer}\label{sec_GEE2_Falconer}
We now derive the GEE2 version of Falconer's method.  Recall from Section \ref{sec_falconer} that Falconer's estimators allow the MZ and DZ population variance parameters to differ.  Thus in deriving GEE2-Falconer, we assume a covariance matrix with two distinct parameters for MZ and DZ population variances ($\sigma^2_{MZ}$, $\sigma_{DZ}^2$) and two distinct correlation parameters ($\rho_{MZ}, \rho_{DZ})$. Thus this approach provides a more flexible way of estimating heritability compared to NACE model which requires the MZ and DZ variance parameters to be the same.  Then define the following quantities which will allow us to fit Falconer's method within the same GEE2-framework presented in Section \ref{sec_GEE2_NACE}: 

\begin{align*}
& Cov(\bs{y}_z)=
\begin{bmatrix} 
\sigma^2_z & \sigma^2_z \rho_z \\  
\sigma^2_z \rho_z & \sigma^2_z
\end{bmatrix} \tab (\text{population trait covariance, } z=\text{MZ or DZ}) \\ 
& g(\sigma^2_z)= v_0 + v_1z \tab  (g \text{ is identity or log-link}) \\
& h(\rho_z) = p_0 + p_1z \tab  (h \text{ is identity or Fisher's Z-transformation}) \\
& \bs{\alpha}=(v_0,v_1,p_0,p_1) \tab (\text{parameters to estimate}) \\
& \bs{\Gamma}_z = (\sigma^2_z,\sigma^2_z,\sigma^2_z \rho_z) \tab \text{(population covariance matrix in vectorized form)}\\
& \bs{\gamma}_z= (y_{z_1}^2,y_{z_2}^2,y_{z_1}y_{z_2}) \tab \text{(sample covariance matrix in vectorized form)} \\
& \bs{\Omega}_z=\bs{\mathrm{I}}_2 \tab (\text{Identity matrix})
\end{align*}

The above implies that $\sigma^2_{MZ}=g^{-1}(v_0+v_1), \ \sigma^2_{DZ}=g^{-1}(v_0), \ \rho_{MZ}=h^{-1}(p_0+p_1), \ \rho_{DZ}=h^{-1}(p_0).$  Equations (\ref{eqn_updater}-\ref{eqn_est_covar}) can be used to obtain $\hat{\bs{\alpha}}$ and $\hat{Cov}(\hat{\bs{\alpha}})$ respectively, which then can be plugged in to get $\hat{\rho}_{MZ},\hat{\rho}_{DZ}$, which then are plugged into Falconer's equations (\ref{eqn_falc}) to get $\hat{h}^2, \hat{c}^2$.  The delta-method is used to obtain the final standard errors and Wald-type confidence intervals for $h^2,c^2$.

Recall that $\bs{\Omega}_z=Cov(\bs{\gamma}_z)$ encodes all assumptions about higher-order moments.  Falconer's estimators only use information from the first two moments thus ignoring all higher-order moments.  Therefore we set $\bs{\Omega}_z=\bs{\mathrm{I}}_2$ so that $\bs{\Omega}$ effectively drops out of equation (\ref{eqn_updater}) which is used to obtain the GEE2-Falconer point estimates.

Lastly, we show how GEE2-Falconer can allow heritability to vary as a function of covariates.  For example, suppose we want to allow heritability to vary as a function of sex.  Then define:
\begin{align*}
& g(\sigma^2_z)= v_0 + v_1z +v_2Sex+v_3 Sex*z \\
& h(\rho_z) = p_0 + p_1z +p_2Sex+p_3 Sex*z \\
\end{align*}
where the new parameters of interest are $\bs{\alpha}=(v_0,v_1,v_2,v_3,p_0,p_1,p_2,p_3).$ Notice that unlike GEE2-NACE, GEE2-Falconer requires covariate-zygosity interactions when allowing $h^2,c^2$ to vary as a function of covariates.  These interaction terms allow the variance and covariance parameters to differ between MZ and DZ twins (we found through simulations that ignoring the interaction terms could lead to under-coverage of the true $h^2$, results not shown).  In contrast, the NACE model assumes all variance components are the same between MZ and DZ twins.   Again, we can use equations (\ref{eqn_updater}-\ref{eqn_est_covar}) to obtain estimates and robust standard errors for $\bs{\alpha}$.
 
Then one can obtain sex-specific estimates of $h^2,c^2$ as follows:
\begin{align*}
& \hat{\rho}_{MZ,Male}=g^{-1}(\hat{p}_0+\hat{p}_1+\hat{p_2}+\hat{p}_3), \ \ \ \hat{\rho}_{DZ,Male}=g^{-1}(\hat{p}_0+\hat{p}_2) \\ &
\hat{\rho}_{MZ,Female}=g^{-1}(\hat{p}_0+\hat{p}_1), \ \ \ \hat{\rho}_{DZ,Female}=g^{-1}(\hat{p}_0) \\ & \hat{h}^2_{Male}= 2(\hat{\rho}_{MZ,Male} - \hat{\rho}_{DZ,Male}), \ \ \ \hat{c}^2_{Male}=2\hat{\rho}_{DZ,Male} - \hat{\rho}_{MZ,Male} \\ & \hat{h}^2_{Female}= 2(\hat{\rho}_{MZ,Female} - \hat{\rho}_{DZ,Female}), \ \ \ \hat{c}^2_{Female}=2\hat{\rho}_{DZ,Female} - \hat{\rho}_{MZ,Female}
\end{align*}
More generally: to estimate the heritability $h^2_{\bs{x}}$ for a particular combination of covariates ``$\bs{x}$'', simply plug $\hat{\rho}_{MZ, \bs{x}},\hat{\rho}_{DZ, \bs{x}}$ into Falconer's 
equations (\ref{eqn_falc}) and use the delta method with $\hat{Cov}(\hat{\bs{\alpha}})$ to get the final standard errors for $\hat{h}^2_{\bs{x}}, \hat{c}^2_{\bs{x}}$.

\section{Results}
\label{sec_results}
In Sections \ref{sec_tACEsim}-\ref{sec_MCTFS}, we compare the following ACE models via simulations and application to real data: the normal ACE model (``NACE''), Falconer's simple moment estimators (``Falconer''), and robust GEE2 versions of both models (``GEE2-NACE'' and ``GEE2-Falconer'' respectively).

\subsection{Estimating Heritability for a Heavy-Tailed Continuous Trait} \label{sec_tACEsim}
\begin{figure}[ht]
\caption{Kernel density of heavy-tailed trait from a randomly selected simulated dataset}
\label{fig_tsim_trait_plot}
\begin{center}
\includegraphics[scale=0.5]{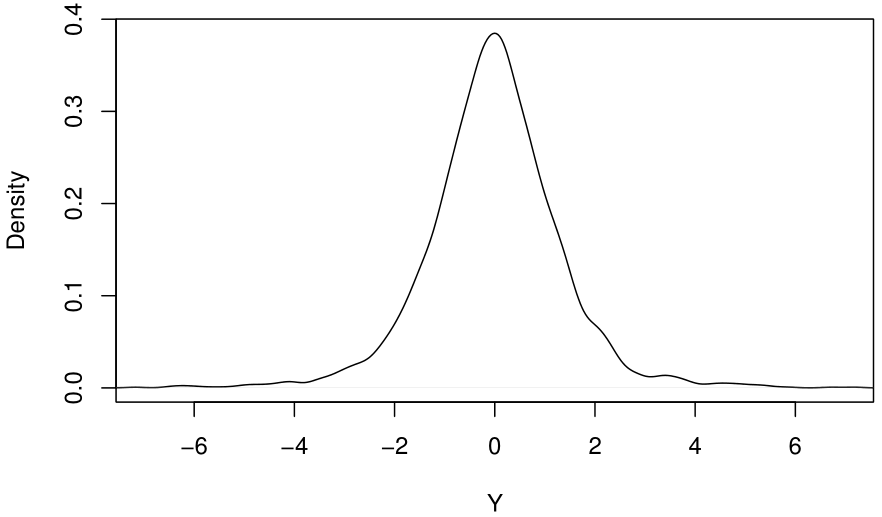}
\end{center}
\end{figure}

Assume the outcome for a given twin pair follows a centered heavy-tailed multivariate t-distribution:
\begin{align} \label{eqn_mvt}
& \bs{y}_z=(y_{z_1}, y_{z_2})\sim f(\bs{y}_z)=\frac{\Gamma(\frac{v+2}{2})}{\Gamma(\frac{v}{2})v\pi|\bs{\Sigma}_z|^{1/2}}\big[1+\frac{1}{v}\bs{y}_z^\T\bs{\Sigma}_z^{-1}\bs{y}_z\big]^{\frac{-(v+2)}{2}} \\ & \bs{\Sigma}_z=\begin{bmatrix}
\sigma^2_A+\sigma^2_C+\sigma^2_E & w_z\sigma^2_A+\sigma^2_C \\ w_z\sigma^2_A+\sigma^2_C & \sigma^2_A+\sigma^2_C+\sigma^2_E
\end{bmatrix} \nonumber
\end{align}
Then with $\sigma^2_A=0.5, \ \sigma^2_C=0.3, \ \sigma^2_E=0.2,$ and $v=4.5$, we simulate 1000 datasets according to (\ref{eqn_mvt}), each with 700 MZ and 700 DZ twin pairs.  See Figure \ref{fig_tsim_trait_plot} for a kernel density plot of the trait from a randomly selected simulated dataset. Among the various models, we are interested in comparing the the following metrics of $h^2$ and $c^2$ across 1000 simulated datasets: the average point estimate, the standard deviation of the estimates (i.e. the ``true standard error''), the average estimated standard error, and the confidence interval coverage rate (i.e. the proportion of all 1000 confidence intervals that contain the true parameter value).

\begin{center}
[Insert Figure \ref{fig_tsim_trait_plot} Here]
\end{center}

From Table \ref{tab_tsimstats}, we see the traditional NACE model has poor coverage for both $h^2$ and $c^2$ (less than $75\%$), whereas GEE2-NACE attains coverage much closer to the nominal rate of $95\%$.  Notice that GEE2-NACE produces identical point estimates to the normal NACE, however, GEE2-NACE produces larger and more trustworthy standard errors.  Table \ref{tab_tsimstats} clearly shows that the average estimated SE's for the NACE significantly underestimate the true SE's; whereas the average estimated SE's for GEE2-NACE match up very well with the true SE's.  The reason the NACE estimated standard errors are incorrect is because they are based on Fisher's Information matrix which is determined by the assumed likelihood function (normal) which is misspecified (the true likelihood is a heavy-tailed t-distribution).  In contrast, GEE2-NACE uses robust sandwich standard errors that provide significantly better coverage of the true variance parameters.

Notice that GEE2-Falconer and Falconer's method produce identical point estimates, however, GEE2-Falconer uses robust standard errors and thus attains significantly better coverage of the true heritability compared to Falconer's method.  A key point is that although the GEE2 models do not attempt to model the true parametric distribution of the trait (heavy-tailed t), they can nevertheless still attain approximately correct coverage rates of the true heritability parameter.

\begin{center}
[Insert Table \ref{tab_tsimstats} here]
\end{center}

\subsection{Estimating Heritability for Right-Skewed Over-Dispersed Count Data}\label{sec_LGPACEsim}
For a given pair of twins, let $\bs{y}_z=(y_{z_1}, y_{z_2})\sim bLGP(\sigma^2_A+\sigma^2_C+\sigma^2_E, \lambda)$, where $bLGP(.)$ is the bivariate Lagrangian Poisson distribution with dispersion parameter $\lambda \in (-1,1)$.  Following Kirkpatrick and Neale\cite{kirkpatrick2016applying}, we can use the \textit{RMKdiscrete} R package \cite{RMKdiscrete} to simulate from the $bLGP$ distribution as follows:\\\\
\begin{minipage}{.5\linewidth}
	For MZ twins:
	\begin{align*}
	& Q_0\sim LGP(\sigma^2_A+\sigma^2_C,\lambda) \\ & Q_1, Q_2 \sim LGP(\sigma^2_E,\lambda) \\ & Y_1= Q_0+Q_1 \text{ and } Y_2=Q_0+Q_2 \\ & \implies Y_1, Y_2 \sim bLGP(\sigma^2_A+\sigma^2_C+\sigma^2_E,\lambda)
	\end{align*}
\end{minipage}%
\begin{minipage}{.5\linewidth}
	For DZ twins:
	\begin{align*}
	& Q_0\sim LGP(0.5\sigma^2_A+\sigma^2_C,\lambda) \\ & Q_1, Q_2 \sim LGP(0.5\sigma^2_A + \sigma^2_E,\lambda) \\ & Y_1= Q_0+Q_1 \text{ and } Y_2=Q_0+Q_2 \\ & \implies Y_1, Y_2 \sim bLGP(\sigma^2_A+\sigma^2_C+\sigma^2_E,\lambda)
	\end{align*}
\end{minipage} \hfill \\
where $LGP(.)$ and $bLGP(.)$ are the univariate and bivariate lagrangian poisson distributions respectively.
Then we have the following distributional properties\cite{kirkpatrick2016applying}: $\mathbb{E}(y_{z_1})=\mathbb{E}(y_{z_2})=\frac{\sigma^2_A+\sigma^2_C+\sigma^2_E}{1-\lambda}$, \ $Var(y_{z_1})=Var(y_{z_1})=\frac{\sigma^2_A+\sigma^2_C+\sigma^2_E}{(1-\lambda)^3}, \ Cov(y_{MZ_1},y_{MZ_2})=\frac{\sigma^2_A+\sigma^2_C}{(1-\lambda)^3}$, \ $Cov(y_{DZ_1},y_{DZ_2})=\frac{0.5\sigma^2_A+\sigma^2_C}{(1-\lambda)^3}$.

\begin{figure}[ht]
\centering
\caption{Histogram of right-skewed over-dispersed count trait from a randomly selected simulated dataset}
\label{fig_lgp_trait_plot}
\begin{center}
\includegraphics[scale=0.5]{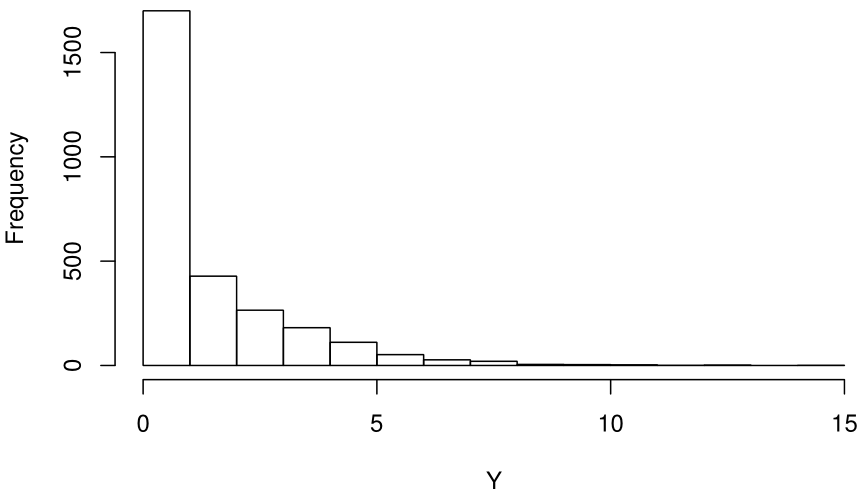}
\end{center}
\end{figure}

However, note that the above construction of the bivariate lagrangian poisson distribution may be invalid when $\lambda<0$ (under-dispersion), but will hold when $\lambda>0$ (over-dispersion)\cite{kirkpatrick2016applying}.  In contrast, our GEE2 ACE models work for both underdispersed or overdispersed count data.  Nevertheless, we will only consider the case of over-dispersed count data with $\lambda=0.35, \sigma^2_A=0.5, \sigma^2_C=0.3,$ and $\sigma^2_E=0.2$. One-thousand datasets are simulated, each with 700 MZ twin pairs and 700 DZ twin pairs.  See Figure \ref{fig_lgp_trait_plot} for a histogram of the trait from a randomly selected simulated dataset.

\begin{center}
[Insert Figure \ref{fig_lgp_trait_plot} here]
\end{center}

Notice from Table \ref{tab_LGPstats} that the same patterns from Section \ref{sec_tACEsim} hold. GEE2-NACE has significantly better coverage rates and more accurate estimated standard errors compared to the traditional NACE.  The same result holds for GEE2-Falconer compared to Falconer's method.  Again, the main problem is that the average estimated standard errors for the NACE and Falconer's method are significantly less than their true standard errors, thus yielding coverage rates much less than the nominal rate of $95\%$.  In contrast, the robust GEE2-NACE and GEE2-Falconer models produce much more accurate standard errors and coverage rates closer to the nominal level.  A key point is that although the GEE2 models do not attempt to model the true parametric distribution of the trait (Lagrangian Poisson), they can nevertheless still attain approximately correct coverage rates of the true heritability parameter.

\begin{center}
[Insert Table \ref{tab_LGPstats} here]
\end{center}

\subsection{Scenario where the NACE Twin model is Biased, but Falconer's Method Remains Unbiased} \label{sec_NACEbiased}
Recall from Section~\ref{sec_falconer} that Falconer's method allows the ACE variance parameters to differ between MZ and DZ twins, as long as the variance \textit{proportions} ($h^2, c^2, e^2$) are the same in MZ and DZ twins. In contrast, the NACE approach makes a stronger assumption that the individual variance components ($\sigma^2_A, \sigma^2_C, \sigma^2_E$) are equal for both MZ and DZ twins.  In the existing literature for the twin NACE model, researchers have made no comments on how to address the scenario where the $\sigma^2_A,\sigma^2_C,\sigma^2_E$ variance components \textit{differ} between MZ and DZ twins\cite{rabe2008biometrical,wang2011statistical,feng2009analysis,mcardle2005mixed}.  Additionally, the assumption of equal variance parameters between MZ and DZ twins is a common criticism of twin studies\cite{rijsdijk2002analytic}.  For example, there is some evidence that MZ twins are treated more similarly by their parents compared to DZ twins\cite{rijsdijk2002analytic}: this may result in MZ twins having smaller shared family environmental variance ($\sigma^2_C$) compared to DZ twins.  Thus it would be beneficial to have methods for estimating heritability that are less sensitive to the assumption of equal variances between MZ and DZ twins (e.g. Falconer's method).

The following simulation study was performed with 700 MZ and 700 DZ twin pairs, where $\bs{y}_z$ follows a bivariate normal distribution with:
\begin{align*}
& Var(y_{MZ_1})=Var(y_{MZ_2}) =   \sigma^2_{A_{MZ}}+\sigma^2_{C_{MZ}}+\sigma^2_{E_{MZ}}=.3+.18+.12=0.6 \\ & Var(y_{DZ_1})=Var(y_{DZ_2}) = \sigma^2_{A_{DZ}}+\sigma^2_{C_{DZ}}+\sigma^2_{E_{DZ}}=.5+.3+.2=1
\end{align*}
 Notice the total MZ variance (0.6) differs from the total DZ variance (1), however, the proportions $h^2=0.5, \ c^2=0.3, \ e^2=0.2$ are equal for both types of twins.

\begin{center}
[Insert Table \ref{tab_nace_bias} here]
\end{center}

Table \ref{tab_nace_bias} shows that NACE produces significantly biased parameter estimates in this setting, while Falconer's method remains approximately unbiased. See Supplemental Figure S1 for an additional demonstration of the NACE bias in this setting.  Therefore, when attempting to fit a twin ACE model, one should first check to see if the total variance is approximately equal for MZ and DZ twins, and if not, Falconer's method (or GEE2-Falconer) should be preferred.

\subsection{Allowing Heritability to vary as a Function of Sex} \label{sec_ACEsex_sim}
Here the ACE variance components are allowed to differ by sex.  Following the notation and assumptions of Sections \ref{sec_normACE} and \ref{sec_GEE2_NACE}, let $a_0=0.3, \ a_1=0.3, \ c_0=0.4, \ c_1= -0.2, \ e_0=0.3,$ and $e_1=-0.1.$  This implies that for males: $\sigma^2_A=0.6, \ \sigma^2_C=0.2, \ E=0.2$ and for females: $\sigma^2_A=0.3, \ \sigma^2_C=0.4, \ \sigma^2_E=0.3.$

For each dataset there are 450 male MZ pairs, 450 female MZ pairs, 450 male DZ pairs, 450 female DZ pairs.  A total of 1000 datasets were simulated.  The results in Table \ref{tab_ACEsexsim} indicate that the average estimated standard errors match up very well with the corresponding true standard errors, thus both models approximately achieve the correct coverage rates for the sex-specific heritability parameters.

\begin{center}
[Insert Table \ref{tab_ACEsexsim} here]
\end{center}

\subsection{Minnesota Center for Twins and Family Study (MCTFS)}\label{sec_MCTFS}
The Minnesota Center for Twins and Family Study (MCTFS)\cite{miller2012minnesota,mcgue2013genome} contains 8,405 subjects clustered into 4-member families (each with 2 parents and 2 twins, either MZ or DZ). The overall goal of the study is to explore the genetic and environmental factors of substance abuse disorders.  We consider five composite quantitative clinical phenotypes\cite{mcgue2013genome}, which were derived using a hierarchical factor analytic approach\cite{hicks2011psychometric}. These five phenotypes are:  1) Nicotine (NIC): composite measure of nicotine use and dependence, 2) Alcohol Consumption (CON): composite of measures of alcohol use frequency and quantity, 3) Illicit Drugs (DRG): composite of frequency of use of 11 different drug classes and DSM symptoms of drug dependence, 4) Behavioral Disinhibition (BD): composite of measures non-substance use behavioral disinhibition including symptoms of conduct disorder and
aggression, and 5) Externalizing Factor (EXT): a composite measure of all five previous traits.  

\begin{figure}[ht]
\caption{Histograms of 5 substance-abuse traits from the Minnesota Center for Twins and Family Study}
\begin{center}
\includegraphics[scale=0.5]{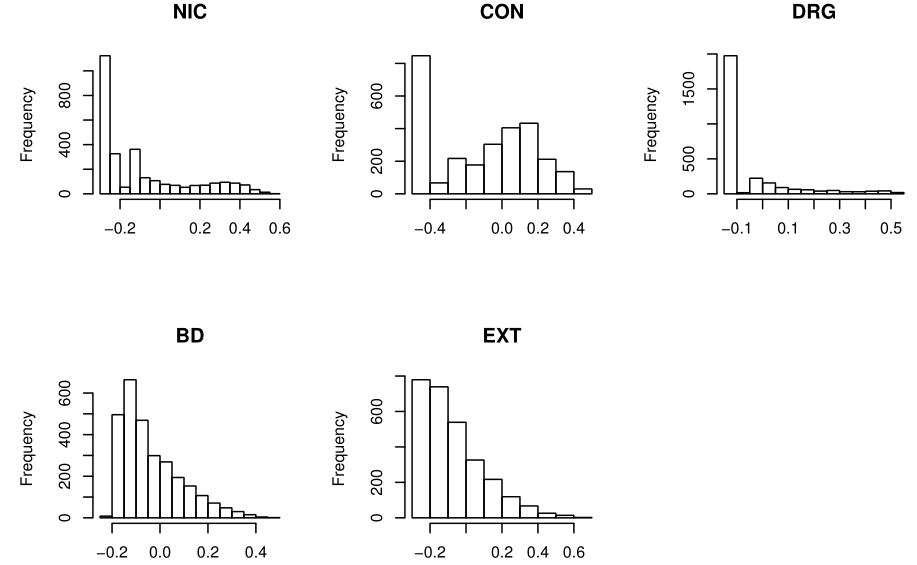}
\end{center}
	\label{fig_RDAhist}
    \caption*{Nicotine (NIC): composite measure of nicotine use and dependence; Alcohol
Consumption (CON): composite of measures of alcohol use frequency and quantity; Illicit
Drugs (DRG): composite of frequency of use of 11 different drug classes and DSM symptoms
of drug dependence; Behavioral Disinhibition (BD): composite of measures non-substance
use behavioral disinhibition including symptoms of conduct disorder and aggression;
Externalizing Factor (EXT): a composite measure of all five previous traits}
\end{figure}

We considered a total of 936 MZ and 478 DZ twin pairs for each phenotype (all twins with non-missing phenotype data, parent data was not included). See Figure \ref{fig_RDAhist} for the histograms of each phenotype; notice that all five phenotypes appear very right-skewed, non-normal, and do not appear to follow any standard parametric distributions.  However, as long as the trait can be approximated by a member of the quadratic exponential family with the first two moments correctly specified, then it is not necessary to try and model the true parametric distribution of these traits, rather one can simply use GEE2 which produces a robust confidence interval of heritability.  Lastly, for all traits, first an ordinary linear model was fit to regress out the effects of several covariates: Sex, Age, and the top 5 principle components; then the residuals were used as the new response for fitting the ACE models.  Although the NACE and GEE2 models can directly adjust for covariate effects, the original Falconer's method cannot.  Thus in order to present a fair comparison between all models, the trait covariate-adjusted residuals were used as the outcome for all models.

The results from Table \ref{tab_RDAstats} indicate several patterns.  First, notice that GEE2-NACE and NACE model produce identical point estimates, however, GEE2-NACE produces larger and probably more trustworthy standard errors (as shown throughout all of simulations).  Similarly, GEE2-Falconer and Falconer's method produce identical point estimates, although the standard errors for GEE2-Falconer are likely more accurate (as shown throughout all simulations).  Interestingly, Falconer's method (and GEE2-Falconer) consistently produce smaller estimates of heritability compared to NACE (and GEE2-NACE).  Recall that the NACE model assumes the population variances are equal between MZ and DZ twins, whereas Falconer's method allows them to differ. Note that the ratio of the MZ to DZ sample variance for the five substance abuse traits is 0.95, 0.99, 0.89, 0.97, and 0.96 respectively.  The fact that the observed sample variances differ between MZ and DZ twins (by at most 11\%) may explain why the NACE and Falconer's method produce different point estimates of heritability in Table \ref{tab_RDAstats} (with a maximum difference of $8\%$ for DRG).

\begin{center}
[Insert Figure \ref{fig_RDAhist} here]
\end{center}

\begin{center}
[Insert Table \ref{tab_RDAstats} here]
\end{center}

\subsubsection{Allow $\bs{h^2,c^2,e^2}$ to vary as a Non-linear Function of Age}\label{sec_MCTFS_age}
The MCTFS is a longitudinal study in which data was collected from a cohort of twins at  five different time periods: ages 11, 17, 20, 24, and 29.  The five quantitative phenotypes in Table \ref{tab_RDAstats} were only available at age 17, however, additional phenotypes related to ``alcoholism'' were available at multiple time points (but not all time periods).  The GEE2-Falconer model was used to jointly model the $h^2,c^2,e^2$ parameters from ages 17-29 for a count phenotype measure of alcohol use (values range from 0 to 5, larger values indicate greater alcohol use). See Supplemental Figure S2 for a histogram of the longitudinal alcohol phenotype.  The GEE2-Falconer model was fit as described in Section \ref{sec_GEE2_Falconer}, with the following modification to allow the $h^2,c^2,e^2$ parameters to vary as a 2nd-degree polynomial function of age:

\begin{align}\label{eqn_FalcACEage}
& g(\sigma^2_z)=v_0 +v_1z+v_2Age+v_3Age^2+v_4Age*z+v_5Age^2*z \nonumber \\
& h(\rho_z)=p_0 +p_1z+p_2Age+p_3Age^2+p_4Age*z+p_5Age^2*z
\end{align}
where $Age$ is the age of a given twin pair, and $Age^2= \big(Age-mean(Age)\big)^2$ is the squared centered age of a given twin pair.  Recall from Section \ref{sec_GEE2_Falconer} that covariate-zygosity interaction terms are necessary when incorporating ACE covariate effects for GEE2-Falconer. The interaction terms allow the correlations and ACE covariate effects to differ between MZ and DZ twins.
Higher-order polynomial effects were not significant (p-values $>0.05$).  For example, to estimate the heritability at age 17, the relevant covariate values are plugged into equation (\ref{eqn_FalcACEage}) to get $\hat{\rho}_{MZ_{17}},\hat{\rho}_{DZ_{17}}$, then $\hat{h}^2_{17}=2(\hat{\rho}_{MZ_{17}}-\hat{\rho}_{DZ_{17}})$.  The Delta-method is used to obtain the relevant standard errors.

\begin{figure}[ht]
\caption{GEE2-Falconer model with $h^2,c^2,e^2$ allowed to vary as a non-linear function of Age (with 95\% confidence intervals) for a longitudinal alcohol use trait from the Minnesota Center for Twins and Family Study}\label{fig_AMT17202429_h2c2e2}
\begin{center}
\includegraphics[scale=0.4]{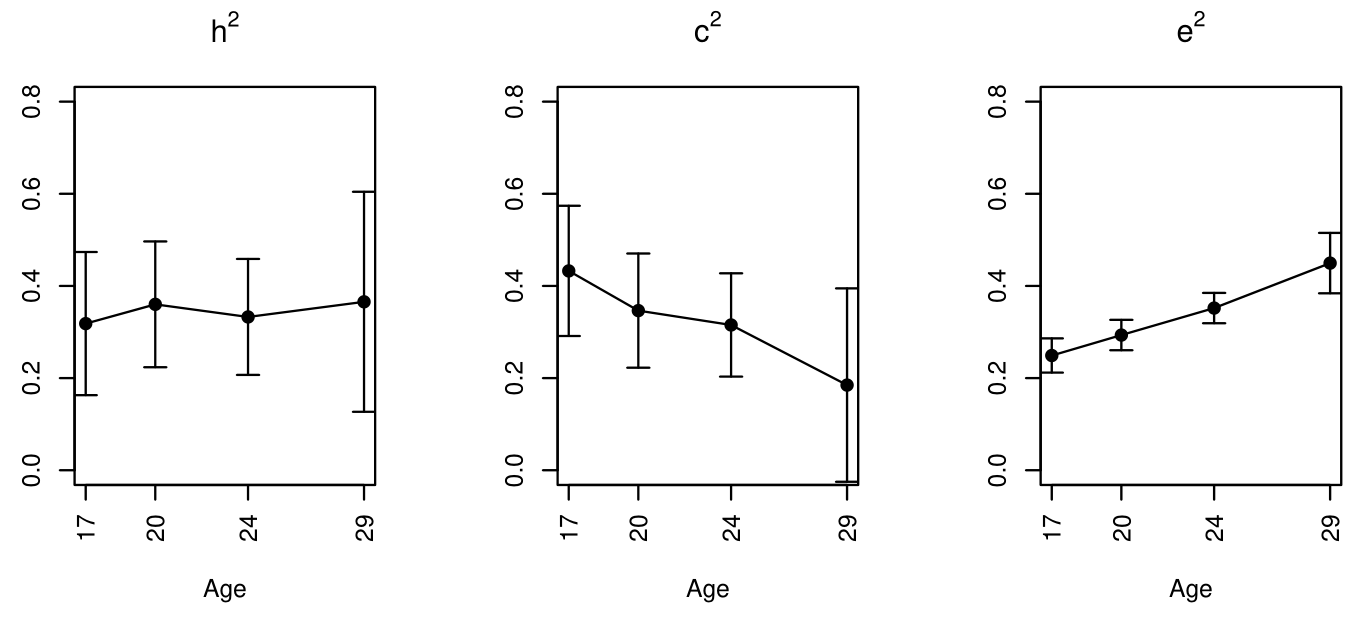}
\end{center}
\caption*{$h^2, c^2, e^2$: proportion of total trait variance due to additive genetic effects, common shared environmental effects, and unique non-shared environmental effects respectively}
\end{figure}

\begin{center}
[Insert Figure \ref{fig_AMT17202429_h2c2e2} here]
\end{center}

Notice from Figure \ref{fig_AMT17202429_h2c2e2} that the non-shared environmental effect ($e^2$) increases over time, while the shared environmental effect ($c^2$) decreases.  The genetic effect ($h^2$) on the Alcohol Use trait remained relatively stable across the four time periods.  Wald tests were used to check if $h^2,c^2,e^2$ significantly changed from ages 17 to 29 (e.g. $\mathrm{H}_0: h^2_{29}-h^2_{17}=0)$ and produced the following p-values respectively: 0.78, 0.092, and $<0.0001$. Intuitively, these results may mean that as the twins age and become more independent, their non-shared environmental experiences have a greater influence on their alcohol use, whereas the effect of their shared-family environment decreases.  Lastly, we note that jointly modeling the Alcohol Use trait at all four time periods resulted in smaller standard errors compared to fitting separate univariate GEE2 models at each time period (see Supplemental Table S1).
\section{Discussion}
\label{sec_discussion}
Twin studies have proven to be powerful instruments in quantifying the genetic and environmental factors of complex phenotypes\cite{neale2013twinsbook,polderman2015meta}.  In practice, the normal ACE model (``NACE'')\cite{rabe2008biometrical,wang2011statistical} and Falconer's moment estimators\cite{falconer1975introduction,rijsdijk2002analytic} are popular methods for estimating heritability in twin studies. We've shown that both models can be fit within a unified second-order generalized estimating equations framework (``GEE2''), which provides robust standard errors and can incorporate covariate effects for both mean and variance parameters (e.g. let heritability vary by sex or age as done in Sections \ref{sec_ACEsex_sim} and \ref{sec_MCTFS_age}).  It's worth emphasizing that the original version of Falconer's method\cite{falconer1975introduction} cannot directly adjust for covariate effects, whereas our GEE2-Falconer model can.

Researchers are often interested in estimating heritability for non-normal phenotypes (e.g. counts, binary, skewed or heavy-tailed continuous data).  When interested in fitting an ACE model to a non-normal phenotype, one option is to try and parametrically model the true distribution\cite{kirkpatrick2016applying,burton_binary,scurrah_poisson,jamsen2013specification}.  However, inferences on the variance components may be sensitive to departures from parametric distributional assumptions.  Our simulations indicate that when the parametric distributional assumption is incorrect, Wald-type confidence intervals for the ACE variance parameters may significantly differ from the nominal rate.  In addition, we've shown that as long as the trait can be approximated by a member of the quadratic exponential family, then it is not necessary to try and fit the true parametric distribution; rather one can simply use GEE2 which provides a robust confidence interval for the true heritability.  The GEE2 model requires only the first two moments
(i.e. mean and variance structures) to be correct, all other moments are allowed to be misspecified.  In contrast, parametric models assume all moments (i.e. the likelihood function) are correct, and may lead to poor coverage rates when assumptions fail.

In Section~\ref{sec_NACEbiased}, we demonstrated an important scenario where NACE produces biased estimates of heritability, while Falconer's method remains unbiased.  Specifically, the NACE assumes that the ACE variance components are \textit{equal} for both MZ and DZ twins (e.g. $\sigma^2_{A_{MZ}}=\sigma^2_{A_{DZ}}$); whereas Falconer's method allows the variance components to differ between twins, and only assumes the variance \textit{proportions} are the same for both twin types (e.g. $h^2_{MZ}=h^2_{DZ}$).  A recent meta-analysis \cite{polderman2015meta} of all twin studies performed in the last 50 years demonstrated that NACE and Falconer's methods can produce substantially different estimates of heritability in practice (see their Supp. Figures 9-10 and Supp. Section 5.7).  Our results highlight one possible explanation for these differences: if the magnitude of the ACE variance parameters differs between MZ and DZ twins (a common criticism of twin studies\cite{rijsdijk2002analytic}), then the NACE will produce biased estimates of heritability, while Falconer's method remains unbiased under weaker assumptions.  Therefore, in practice, Falconer's method should be preferred. 

Although this paper focuses on the ACE model, all models considered can be extended to fit the ``ADE'' twin model, where ``D'' stands for genetic dominance effects.  In practice, researchers typically fit an ACE model if $r_{DZ}>0.5r_{MZ}$, and an ADE model when $r_{DZ}<0.5r_{MZ}$\cite{rijsdijk2002analytic}.  However, we chose to focus on the ACE model for several reasons: 1) both\cite{zaitlen2013using,tenesa2013heritability} found that ignoring shared environmental effects lead to greater bias in estimated heritability compared to ignoring dominance or epistatic genetic effects.  2) Assuming the true model is ACDE, Wang et al\cite{wang2011statistical} proved that $\hat{\sigma}^2_{A}$ from a working ACE model is a consistent estimator of $\sigma^2_{A}+1.5\sigma^2_D$; while $\hat{\sigma}^2_{A}$ from a working ADE model is a consistent estimator of $\sigma^2_{A}+3\sigma^2_C$.  Notice the working ACE model estimate of $\sigma^2_A$ only reflects genetic effects (both additive and dominant), while the working ADE model estimate of $\sigma^2_A$ is confounded/biased by shared environmental effects.  Thus if the goal is to estimate heritability (the proportion of trait variance due to genetic effects), then the working ACE model seems preferable to the working ADE model under model misspecification.   3) Our real data application focused on substance abuse disorder traits, which have been shown to have substantial shared family environmental effects\cite{polderman2015meta}.

In summary, we've shown that given non-normal data, the traditional normal NACE or Falconer's method may significantly undercover the true heritability parameter.  In contrast, the proposed GEE2 models can obtain valid inference for the heritability of a wide variety of data types, such as: normal, binary, counts, heavy-tailed or skewed data. The GEE2 framework requires only the first two moments (i.e. mean and variance structures) to be correctly specified, while all higher-order moments are allowed to be modeled incorrectly.  We showed that both the traditional NACE and Falconer's methods can be fit within a unified GEE2 framework which provides robust standard errors and can incorporate covariate effects in mean and variance-level parameters (e.g. let heritability vary as a function of age or sex).  It is important to note that the traditional Falconer's method\cite{falconer1975introduction} cannot directly adjust for covariate effects whereas our GEE2-Falconer model can.  Finally, we demonstrated that if the ACE variance parameters differ between MZ and DZ twins, then the standard NACE produces biased estimates of heritability, while Falconer's method can still produce unbiased estimates in such settings. Overall, we recommend using the robust and flexible GEE2-Falconer model
for estimating heritability in twin studies.

\section*{Supplemental Data}
Supplemental Data Section 1 includes two figures and one table. Section 2 shows that the NACE and GEE2-NACE estimating equations are identical.

\section*{Declaration of Interests}
The authors declare no competing interests.
\section*{Acknowledgments}
This research was supported by the NIH grant R01DA033958 (PI: Saonli Basu) and NIH grant T32GM108557 (PI: Wei
Pan).
\section*{Web Resources}
R code for fitting all models considered in this paper will be available at \url{https://github.com/arbet003}.

\bibliographystyle{ajhg}
\bibliography{bib}
\clearpage
\section*{Figure Titles and Legends}
\FloatBarrier

\FloatBarrier

\FloatBarrier

\FloatBarrier

\section*{Tables}

\begin{table}[ht]
\centering
\caption{Heavy-tailed trait simulation: mean point estimates ($\bar{h}^2, \bar{c}^2)$, true standard errors ``$SE$'' (standard deviation of estimates across all simulated datasets), mean estimated standard errors ($\bar{SE}$), and $95\%$ confidence interval coverage rates of $h^2=0.5$ and $c^2=0.3$ across 1000 simulated datasets}
\label{tab_tsimstats}
\begin{tabular}{llll}
\hline
Model & $\bar{h}^2 (SE, \ \bar{SE})$  & $\bar{c}^2 (SE, \ \bar{SE})$ & Coverage ($h^2,c^2$)\\ 
\hline
NACE & 0.50 (0.10, 0.05) & 0.30 (0.09, 0.05) & (0.74, 0.74) \\
GEE2-NACE & 0.50 (0.10, 0.09) & 0.30 (0.09, 0.08) & (0.95, 0.94) \\
Falconer & 0.50 (0.10, 0.04) & 0.30 (0.09, 0.04) & (0.58, 0.60) \\
GEE2-Falconer & 0.50 (0.10, 0.10) & 0.30 (0.09, 0.09) & (0.95, 0.95) \\
\hline
\end{tabular}
\end{table}
\FloatBarrier

\begin{table}[ht]
\centering
\caption{Right-skewed over-dispersed count trait simulation: mean point estimates ($\bar{h}^2, \bar{c}^2)$, true standard errors ``$SE$'' (standard deviation of estimates across all simulated datasets), mean estimated standard errors ($\bar{SE}$), and $95\%$ confidence interval coverage rates of $h^2=0.5$ and $c^2=0.3$ across 1000 simulated datasets}
\label{tab_LGPstats}
\begin{tabular}{llll}
\hline
Model & $\bar{h}^2 (SE, \ \bar{SE})$  & $\bar{c}^2 (SE, \ \bar{SE})$ & Coverage ($h^2,c^2$)\\ 
\hline
NACE & 0.50 (0.11, 0.05) & 0.30 (0.10, 0.05) & (0.63, 0.67) \\
GEE2-NACE & 0.50 (0.11, 0.11) & 0.30 (0.10, 0.10) & (0.95, 0.94) \\
Falconer & 0.50 (0.11, 0.04) & 0.30 (0.10, 0.04) & (0.54, 0.55) \\
GEE2-Falconer & 0.50 (0.11, 0.12) & 0.30 (0.10, 0.10) & (0.95, 0.94) \\
\hline
\end{tabular}
\end{table}
\FloatBarrier

\begin{table}[ht]
\centering
\caption{Scenario where NACE is biased and Falconer's method is unbiased: Average point estimates ($\bar{h}^2, \bar{c}^2$) across 1000 simulated datasets (standard error of mean ``SEM'' in parentheses)}
\label{tab_nace_bias}
\begin{tabular}{lll}
\hline
  & $\bar{h}^2$ & $\bar{c}^2$ \\ 
\hline
Truth & 0.50 & 0.30 \\
Falconer & 0.50 (0.002) & 0.30 (0.002) \\
NACE & 0.70 (0.002) & 0.15 (0.002) \\
\hline
\end{tabular}
\end{table}
\FloatBarrier

\begin{table}[ht]
\centering
\caption{Simulation allowing heritability ($h^2$) to vary by sex: average point estimates ($\bar{h}^2_{Male},\bar{h}^2_{Female}$) across 1000 simulated datasets.  In parentheses: true standard error (standard deviation of estimates across all datasets), average estimated standard error, and 95\% confidence interval coverage rate}
\label{tab_ACEsexsim}
\begin{tabular}{lll}
\hline
Model & $\bar{h}^2_{Male}$  & $\bar{h}^2_{Female}$ \\ 
\hline
Truth & 0.60 & 0.30 \\
GEE2-NACE & 0.60 (0.07, 0.07, 0.96) & 0.30 (0.07, 0.07, 0.94) \\
GEE2-Falconer & 0.60 (0.08, 0.08, 0.96) & 0.30 (0.08, 0.08, 0.95) \\
\hline
\end{tabular}
\end{table}
\FloatBarrier

\begin{table}[ht]
\centering
\caption{Real data analysis point estimates and standard errors (in parentheses) for 5 substance-abuse traits from the Minnesota Center for Twins and Family Study}
\label{tab_RDAstats}
\begin{tabular}{llll}
\hline
Trait & Model & $h^2$ & $c^2$ \\ 
\hline
NIC & NACE & 0.53 (0.07) & 0.19 (0.07) \\
 & GEE2-NACE & 0.53 (0.10) & 0.19 (0.09) \\
 & Falconer & 0.48 (0.05) & 0.24 (0.05) \\
 & GEE2-Falconer & 0.49 (0.10) & 0.23 (0.09) \\
\hline
CON & NACE & 0.44 (0.06) & 0.29 (0.06) \\
 & GEE2-NACE & 0.44 (0.09) & 0.29 (0.08) \\
 & Falconer & 0.40 (0.05) & 0.32 (0.05) \\
 & GEE2-Falconer & 0.40 (0.09) & 0.32 (0.08) \\
\hline
DRG & NACE & 0.50 (0.07) & 0.20 (0.07) \\
 & GEE2-NACE & 0.50 (0.13) & 0.20 (0.12) \\
 & Falconer & 0.42 (0.06) & 0.26 (0.05) \\
 & GEE2-Falconer & 0.42 (0.12) & 0.26 (0.11) \\
\hline
BD & NACE & 0.67 (0.07) & 0.08 (0.07) \\
 & GEE2-NACE & 0.67 (0.09) & 0.08 (0.09) \\
 & Falconer & 0.63 (0.06) & 0.12 (0.05) \\
 & GEE2-Falconer & 0.63 (0.09) & 0.12 (0.09) \\
\hline
EXT & NACE & 0.60 (0.06) & 0.18 (0.06) \\
 & GEE2-NACE & 0.60 (0.09) & 0.18 (0.09) \\
 & Falconer & 0.55 (0.05) & 0.23 (0.05) \\
 & GEE2-Falconer & 0.55 (0.09) & 0.22 (0.08) \\
\hline
\end{tabular}
\caption*{Nicotine (NIC): composite measure of nicotine use and dependence; Alcohol
Consumption (CON): composite of measures of alcohol use frequency and quantity; Illicit
Drugs (DRG): composite of frequency of use of 11 different drug classes and DSM symptoms
of drug dependence; Behavioral Disinhibition (BD): composite of measures non-substance
use behavioral disinhibition including symptoms of conduct disorder and aggression;
Externalizing Factor (EXT): a composite measure of all five previous traits}
\end{table}
\FloatBarrier
\clearpage
\end{document}